\documentclass{article}

\usepackage{arxiv}

\usepackage[utf8]{inputenc} 
\usepackage[T1]{fontenc}    
\usepackage{nicefrac}       
\usepackage{microtype}      
\usepackage{amsmath}
\usepackage{amssymb}
\usepackage{amsfonts}
\usepackage{amsthm}         

\usepackage{graphicx}
\usepackage{booktabs}       
\usepackage{multirow}
\usepackage[table]{xcolor}
\usepackage[caption=false,font=normalsize,labelfont=sf,textfont=sf]{subfig}

\usepackage[ruled,vlined,linesnumbered]{algorithm2e}
\SetKwInput{KwInput}{Input}
\SetKwInput{KwOutput}{Output}
\SetKw{KwInit}{Initialize}
\SetKw{KwRet}{return}
\SetKw{KwAnd}{and}
\SetKw{KwOr}{or}

\usepackage{url}          
\usepackage{natbib}
\let\cite\citep
\usepackage{doi}
\usepackage{hyperref}       

\usepackage[shortcuts,acronym]{glossaries}
\glsdisablehyper
\newif\ifuniqueAffiliation
\uniqueAffiliationfalse

\usepackage{authblk}

\newbox{\orcid}\sbox{\orcid}{\includegraphics[scale=0.06]{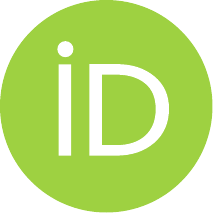}} 

\title{Descriptor: Multi-Regional Cloud Honeynet Dataset (MURHCAD) \thanks{This work has been submitted to the IEEE for possible publication}}

\date{}

\author[1]{%
    \href{https://orcid.org/0009-0005-5068-3166}{\usebox{\orcid}\hspace{1mm}Enrique Feito-Casares}%
}
\author[1]{%
    \href{https://orcid.org/0000-0003-4673-8193}{\usebox{\orcid}\hspace{1mm}Ismael Gómez-Talal\thanks{\texttt{ismael.gomez.talal@urjc.es}}}%
}
\author[1,2,3]{%
    \href{https://orcid.org/0000-0003-0426-8912}{\usebox{\orcid}\hspace{1mm}José-Luis Rojo-Álvarez}%
}

\affil[1]{Department of Signal Theory and Communications and Telematic Systems and Computation, \protect\\ Universidad Rey Juan Carlos, Fuenlabrada, Madrid, Spain}
\affil[2]{D!lemmaLab Ldt, Fuenlabrada, Madrid, Spain}
\affil[3]{Centro de Investigación for Data, Complex Networks and Cybersecurity Sciences,  \protect\\ Universidad Rey Juan Carlos, Madrid, Spain.}

\renewcommand{\shorttitle}{Multi-Regional Cloud Honeynet Dataset}

\hypersetup{
pdftitle={Descriptor: Multi-Regional Cloud Honeynet Dataset (MURHCAD)},
pdfsubject={cs.CR, cs.NI},
pdfauthor={Gómez-Talal et al.},
pdfkeywords={Anomaly Detection, Azure Cloud Deployment, Cyberattack Dataset, Honeynet, Temporal Attack Patterns},
}

\begin{document}
\maketitle

\begin{abstract}
This data article introduces a comprehensive, high-resolution honeynet dataset designed to support standalone analyses of global cyberattack behaviors. Collected over a continuous 72-hour window (June 9 to 11, 2025) on Microsoft Azure, the dataset comprises 132,425 individual attack events captured by three honeypots (Cowrie, Dionaea, and SentryPeer) deployed across four geographically dispersed virtual machines. Each event record includes enriched metadata (UTC timestamps, source/destination IPs, autonomous system and organizational mappings, geolocation coordinates, targeted ports, and honeypot identifiers alongside derived temporal features and standardized protocol classifications). We provide actionable guidance for researchers seeking to leverage this dataset in anomaly detection, protocol-misuse studies, threat intelligence, and defensive policy design. Descriptive statistics highlight significant skew: 2,438 unique source IPs span 95 countries, yet the top 1\% of IPs account for 15\%  of all events, and three protocols dominate: Session Initiation Protocol (SIP), Telnet, Server Message Block (SMB). Temporal analysis uncovers pronounced rush-hour peaks at 07:00 and 23:00 UTC, interspersed with maintenance-induced gaps that reveal operational blind spots. Geospatial mapping further underscores platform-specific biases: SentryPeer captures concentrated SIP floods in North America and Southeast Asia, Cowrie logs Telnet/SSH scans predominantly from Western Europe and the U.S., and Dionaea records SMB exploits around European nodes. By combining fine-grained temporal resolution with rich, contextual geolocation and protocol metadata, this standalone dataset aims to empower reproducible, cloud-scale investigations into evolving cyber threats. Accompanying analysis code and data access details are provided to facilitate immediate adoption and foster further methodological innovations in cybersecurity research.
\end{abstract}

\keywords{Anomaly Detection \and Azure Cloud Deployment \and Cyberattack Dataset \and Honeynet \and Temporal Attack Patterns}


\section*{BACKGROUND}
The design and evaluation of effective cybersecurity solutions strongly depend on the availability of realistic and diverse datasets. Traditional datasets have frequently been limited either in scope or realism, often lacking real-time attack traces or contextual metadata. Recent contributions aim to address this gap by providing curated traffic traces, logs, and behavioral patterns in controlled but realistic environments. For example, a dataset was proposed for capturing comprehensive attack scenarios against vulnerable virtual machines, including FTP and SSH exploitation, ARP spoofing, and brute-force attempts, all observed via tools such as Wireshark and \texttt{tcpdump} ~\cite{shandilya2023cyber} . 
Cyber-physical systems have been explored through labeled datasets with breakdowns, anomalies, and cyberattacks on fluid control subsystems, providing valuable insight into security monitoring in industrial contexts~\cite{laso2017dataset}. 

Dynamic and large-scale initiatives have contributed realistic network traffic and system logs obtained during multi-day cyber defense exercises~\cite{tovarvnak2020traffic}, allowing researchers to test anomaly detection methods under simulated attack–defense scenarios. At the TCP level, buffering delays in cyber-physical systems and Internet of Things (IoT) environments have been characterized, offering fine-grained data on how protocol-level interactions can degrade responsiveness in real-time systems~\cite{al2021dataset}. In the IoT domain, where device heterogeneity and Distributed Denial of Service (DDoS) threats pose unique challenges, recent datasets focus on classifying and detecting diverse attack types in heterogeneous networks, such as IoT-DH~\cite{saif2024iot}. Large-scale industrial data from optical–wireless sensor networks has also been made available, enabling cyber-physical systems performance analysis with metrics such as latency, throughput, and packet error rates~\cite{faheem2022big}. More relevant to our work are honeypot-based initiatives. Among these, Hornet 40~\cite{valeros2022hornet} and Hornet 65 Niner~\cite{valeros2025ctu} emphasize the role of geographical location in attack exposure, aggregating millions of flow records from passive sensors and honeypots deployed globally (e.g., New York, London, Bangalore), and providing standardized Zeek or Argus logs.

To address this gap, we propose the MURHCAD , a high-resolution collection of inbound cyberattacks captured over a 72-hour period via a globally distributed honeynet on Microsoft Azure. This dataset facilitates novel investigations into attack bursts, diurnal rhythms, protocol-level misuse, and concentration of malicious behavior by source IPs and organizations.  In addition to canonical attributes such as timestamp, protocol, and port information, the dataset enriches each event with source geolocation, organizational attribution (via IP-to-autonomous system number (ASN) mappings), and the targeted host. Unlike prior honeypot collections, this dataset places particular emphasis on temporal granularity-allowing researchers to investigate attack behavior patterns at the hourly level-and supports analysis of protocol misuse, attack bursts, and spatial-temporal attack dynamics.

While previous honeypot datasets have offered valuable insights into global attack surfaces, they often lack high-resolution temporal labeling, detailed metadata on attacker provenance, and synchronized data across heterogeneous platforms. In particular, the literature still lacks a dataset that combines large-scale, geographically distributed honeypot traces with hourly-level resolution and enriched contextual metadata, enabling fine-grained spatiotemporal analyses ~\cite{ring2019survey}. 

The presented dataset \cite{Feito-Casares2025b} delivers a comprehensive portrait of real-world cyberattacks by combining honeypot logs from three distinct platforms (Cowrie, Dionaea, and SentryPeer\footnote{Registered
trademark.}) deployed across four geographically dispersed virtual machines. A honeypot is a security mechanism designed to attract and study cyberattacks by simulating vulnerable systems. This dataset fine-grained temporal resolution enables researchers to dissect attack dynamics on an hourly basis, revealing characteristic daily rush hours and maintenance-induced blind spots. Simultaneously, protocol-level annotations (e.g., Session Initiation Protocol (SIP), Telnet, or Server Message Block (SMB)) highlight adversaries’ shifting focus among VoIP, legacy remote-login services, and Windows file-sharing endpoints, providing actionable insights for network segmentation and firewall policy design.

Beyond temporal and protocol analyses, the Multi-Regional Cloud Honeynet Dataset offers rich geolocation metadata that supports cross-regional threat intelligence: over 2 400 unique source IPs spanning 95 countries illuminate global scanning patterns and concentrate attention on high-connectivity hubs. By mapping each intrusion to its autonomous system organization and country of origin, one can evaluate the effectiveness of anomaly-detection algorithms in distinguishing benign background noise from orchestrated scanning campaigns and further refine machine-learning classifiers for multi-class, time-series scenarios.

Because this corpus captures both qualitative (attackType, protocol, srcCountry) and quantitative (timestamps, ports, latitude/longitude) dimensions within a consistent three-day window, it is especially well suited for benchmarking defensive mechanisms under controlled yet realistic conditions. Industrial practitioners may leverage these insights to optimize honeypot placement and adaptive monitoring, while academic researchers can use the data to develop and validate generative models, forecasting tools, and automated segmentation strategies that anticipate and mitigate emerging cyber threats.

\section*{COLLECTION METHODS AND DESIGN}
\begin{figure*}[ht]
    \centering
    \includegraphics[width=0.75\linewidth]{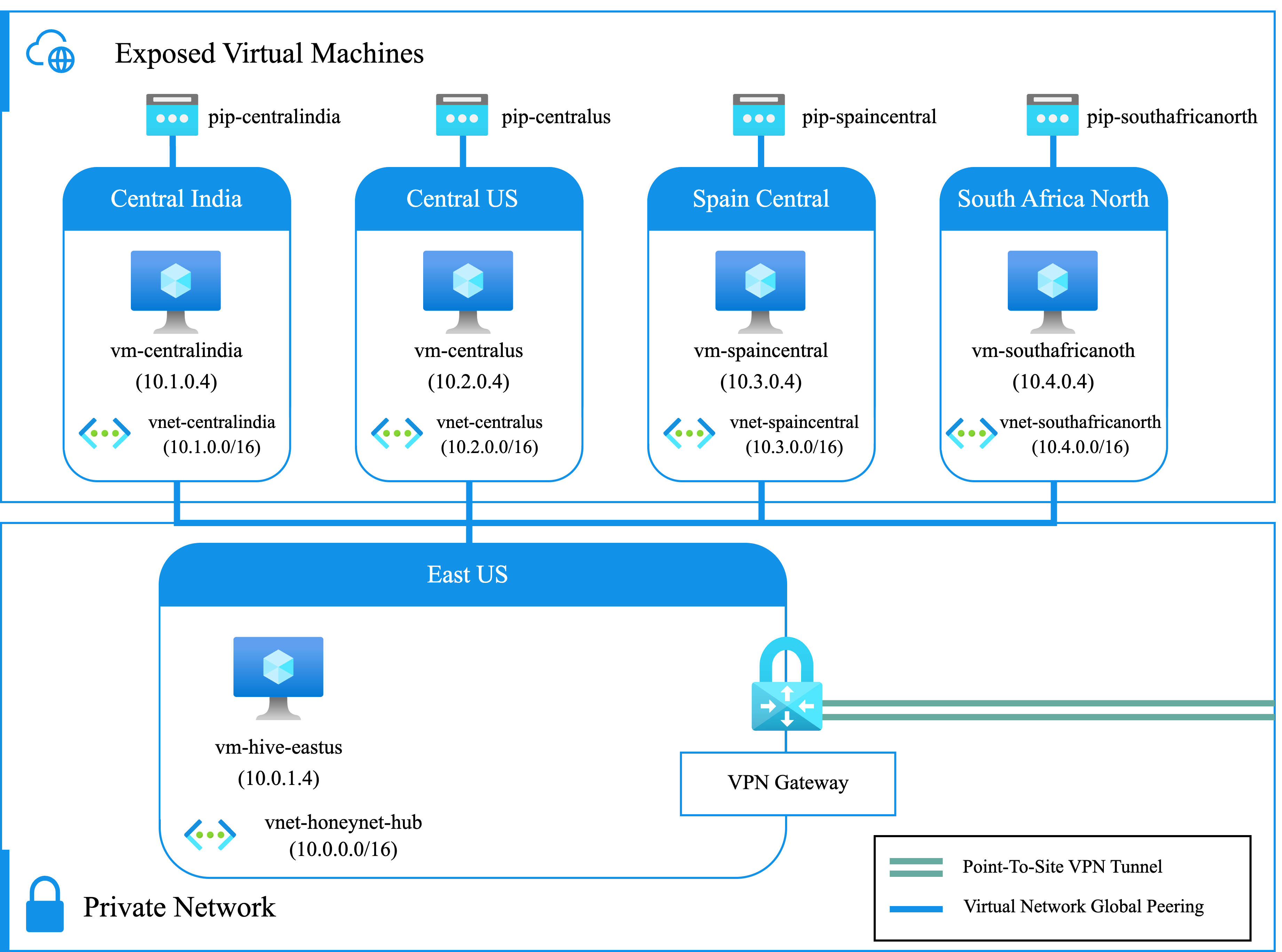}
    \caption{Global Azure-based hub-and-spoke honeynet architecture containing four exposed sensor virtual machines, each within its virtual network, all peered to a central honeynet hub virtual network hosting the hive virtual machine and a VPN Gateway.}
    \label{fig:network-diagram}
\end{figure*}

The experimental protocol involved deploying four virtual machines across different geographic regions using the Azure cloud platform to mimic a realistic environment. Each virtual machine ran honeypot services to monitor and log activity. The virtual machines remained active from June 9 to June 11, enabling comprehensive data collection.

The dataset described in this work was extracted using a dedicated infrastructure on the Microsoft Azure cloud provider and using the T-Pot Community Edition \cite{tpotce}. T-Pot Community Edition is an all-in-one, optionally distributed, multi-architecture honeypot platform supporting over 20 honeypots and offering rich visualization, animated live attack maps, and numerous security tools to enhance the deception experience.

To capture high-value threat activity on key service ports, three distinct honeypot tools were selected and deployed with mirrored configurations:
\begin{itemize}
  \item Cowrie: Emulated SSH and Telnet services with a virtual filesystem, customizable banners, and interactive shell responses. It supports multiple concurrent sessions for high-fidelity brute-force and credential-harvesting capture \cite{cowrie}.
  \item Dionaea: Emulated services such as SMB, MSSQL, MySQL, TFTP, and MQTT to capture malware and exploit payloads through simulated network interactions \cite{dionaea}.
  \item SentryPeer: Monitored SIP traffic for INVITE floods and malformed requests, logging source IP addresses and attempted dialed numbers \cite{Sentrypeer}.
\end{itemize}

The honeynet was implemented as a hub-and-spoke topology within Microsoft Azure. Four sensor virtual networks (VNets) in Central India, Central US, Spain Central, and South Africa North host identical preemptible virtual machines running Cowrie, Dionaea, and SentryPeer honeypots (see Table~\ref{table:sensors}). Figure \ref{fig:network-diagram} shows a detailed diagram of the described architecture. Each sensor VNet was assigned a /16 address space and protected by Network Security Groups that permit inbound traffic only on the specific honeypot service ports. Each sensor virtual machine exposed a public IP interface permitting unrestricted inbound traffic on all honeypot service ports while administrative ports remained inaccessible from the Internet and were only reachable from the VNet. All sensor VNets were peered to a central hive VNet in East US (see Table~\ref{table:hive}). Azure private global backbone was used for peering, providing low-latency connectivity without public Internet exposure. The hive VNet included a Point-to-Site VPN Gateway configured for SSL/TLS tunnels, through which administrators established SSH sessions to both hive and sensor instances over the private network. All telemetry from sensor instances was aggregated in the hive for ingestion, enabling centralized indexing, visualization, and long-term archival.

\begin{table*}
  \caption{Configuration and characteristics of sensor virtual machines.}
  \centering
  \begin{tabular}{@{}lllllll@{}}
    \toprule
    \textbf{Region}  & \textbf{Virtual Machine Name} & \textbf{SKU \& Specs} & \textbf{Disk} & \textbf{OS Image} & \textbf{Private IP} \\
    \midrule
    Central India           &  vm-centralindia     & Standard\_D2as\_v5  & 64 GiB  & Ubuntu 24.04 LTS & 10.1.0.4 \\
    Central US                &  vm-centralus        & Standard\_D2as\_v5  & 64 GiB  & Ubuntu 24.04 LTS & 10.2.0.4 \\
    Spain Central        & vm-spaincentral     & Standard\_D2as\_v5   & 64 GiB  & Ubuntu 24.04 LTS & 10.3.0.4 \\
    South Africa North  &  vm-southafricanorth & Standard\_D2as\_v5   & 64 GiB & Ubuntu 24.04 LTS & 10.4.0.4 \\
    \bottomrule
  \end{tabular}
  \label{table:sensors}
  
\end{table*}

\begin{table*}
 \caption{Configuration and characteristics of the hive virtual machine.}
  \centering
  \begin{tabular}{@{}lllllll@{}}
    \toprule
    \textbf{Region} & \textbf{Virtual Machine Name} & \textbf{SKU \& Specs} & \textbf{Disk} & \textbf{OS Image} & \textbf{Private IP} \\
    \midrule
    East US  & vm-hive-eastus & Standard\_B4ms & 128 GiB  & Ubuntu 24.04 LTS & 10.0.1.4 \\
    \bottomrule
  \end{tabular}
  \label{table:hive}
\end{table*}

In addition to the provided dataset, a full tutorial for deploying the described infrastructure and installing all necessary tools to replicate the experiment is provided in the project repository at \cite{Feito-Casares2025}.

\section*{VALIDATION AND QUALITY}

The dataset underwent comprehensive validation to ensure technical quality, accuracy, and consistency of the collected data. Multiple validation approaches were employed to verify the integrity and reliability of the honeypot infrastructure and data collection pipeline.

\subsection*{Dataset Composition} 
The Multi-Regional Cloud Honeynet Dataset comprises 132,425 rows of logs captured through the honeypot systems, consisting of cyberattacks that occurred between June 9 and June 11, 2025. Each row corresponds to a unique attack interaction recorded, denoting distinct captures by honeypots distributed across several virtual machines in different global locations. To enhance interpretability, Table~\ref{tab:feature_description} provides a concise semantic description of each variable in the dataset. This includes geospatial, temporal, protocol-level, and infrastructure-level indicators captured from the honeypot network. As indicated in Tables~\ref{tab:categorical_summary} and~\ref{tab:numerical_summary}, this dataset exhibits both quantitative and qualitative variables.

\subsection*{Collecting Infrastructure} 
The Azure-based infrastructure maintained high uptime across all virtual machines during the 72-hour collection period. Documented maintenance windows involved planned shutdowns, which created identifiable gaps in data collection (visible as sawtooth patterns in Figure~\ref{fig:temporal_patterns}).

\subsection*{Data Metrics} Comprehensive statistical analysis revealed consistent data quality across all collection periods. The extreme concentration pattern (top 1\% of IPs accounting for 15\% of events) represents genuine attacker behavior rather than data artifacts. Port distribution analysis confirmed realistic patterns with ephemeral source ports and targeted destination ports concentrated on common service endpoints, matching expected attack patterns.

\begin{table*}[ht]
\centering
\caption{Variable description.}
\label{tab:feature_description}
\begin{tabular}{lp{10.5cm}}
\toprule
\textbf{Variable} & \textbf{Description} \\
\midrule
\texttt{weekday} & Day of the week when the event occurred (e.g., \textit{Tuesday}). \\
\texttt{date} & Exact date of the event in MM/DD/YY format. \\
\texttt{srcIp} & Public IP address from which the attack originated. \\
\texttt{srcOrg} & Organization associated with the source IP block (based on ASN lookup). \\
\texttt{srcCountryName} & Country name linked to the source IP. \\
\texttt{srcGeoIp} & Redundant value of \texttt{srcIp}, typically used for geolocation. \\
\texttt{dstIp} & Public IP address of the target system. \\
\texttt{dstIpInternal} & Private IP address of the internal host receiving the attack. \\
\texttt{dstHostname} & Hostname of the machine (virtual machine or container) under attack. \\
\texttt{dstGeoIp} & IP used to derive geolocation of the target (similar to \texttt{dstIp}). \\
\texttt{dstCountryName} & Country in which the targeted system is located (via destination IP). \\
\texttt{protocol} & Protocol used during the attack (e.g., \texttt{sip}, \texttt{telnet}, \texttt{smbd}). \\
\texttt{attackType} & Honeypot system that captured the event: \texttt{Cowrie}, \texttt{Dionaea}, or \texttt{SentryPeer}. \\
\texttt{hour} & Hour of the day (0–23) when the event was logged. \\
\texttt{day} & Numeric day extracted from the timestamp. \\
\texttt{srcPort} & Source port number used by the attacker. \\
\texttt{srcLat}, \texttt{srcLon} & Geographic coordinates (latitude and longitude) of the attack origin. \\
\texttt{dstPort} & Destination port targeted by the attacker. \\
\texttt{dstLat}, \texttt{dstLon} & Geographic coordinates of the targeted system. \\
\bottomrule
\end{tabular}
\end{table*}

\begin{table}[h]
\centering
\caption{Summary of categorical features in the dataset. It includes temporal, geographical, and protocol-level attributes, showing the number of unique values, the most frequent value (Top), and its frequency (Freq) for each feature.}
\label{tab:categorical_summary}
\begin{tabular}{lrrr}
\toprule
\textbf{Feature} & \textbf{Unique} & \textbf{Top} & \textbf{Freq} \\
\midrule
weekday              & 3    & Tuesday        & 76318 \\
date                 & 3    & 6/10/25        & 76318 \\
srcIp                & 2438 & 23.175.48.211  & 18561 \\
srcOrg               & 522  & VDI-NETWORK    & 18561 \\
srcCountryName       & 95   & United States  & 38600 \\
srcGeoIp             & 2438 & 23.175.48.211  & 18561 \\
dstIp                & 11   & 135.235.171.89 & 47370 \\
dstIpInternal        & 4    & 10.1.0.4       & 53057 \\
dstHostname          & 4    & vm-centralindia   & 53057 \\
dstGeoIp             & 11   & 135.235.171.89 & 47370 \\
dstCountryName       & 4    & India          & 53057 \\
protocol             & 13   & sip            & 55060 \\
attackType           & 3    & SentryPeer     & 55060 \\
\bottomrule
\end{tabular}
\end{table}

\begin{table}[h]
\centering
\small
\caption{Summary statistics of the numerical features in the dataset. This includes temporal indicators (e.g., hour and day), source and destination port numbers, and geographical coordinates (latitude and longitude) for both source and destination IPs.}
\label{tab:numerical_summary}
\begin{tabular}{lrrrrrr}
\toprule
\textbf{Feature} & \textbf{Mean} & \textbf{Std} & \textbf{25\%} & \textbf{50\%} & \textbf{75\%} \\
\midrule
hour             & 11.41 & 7.19   & 6      & 11     & 18 \\
day              & 10.18 & 0.60  & 10     & 10     & 11 \\
srcPort          & 48603.54 & 15328.04  & 38833.75 & 55630 & 58804 \\
srcLat           & 30.33 & 18.47  & 22.26  & 37.75  & 37.75 \\
srcLon           & 19.53 & 86.91  & -97.82 & 25.58  & 103.97 \\
dstPort          & 2693.48 & 3933.13  & 135    & 445    & 5060 \\
dstLat           & 25.01 & 20.12   & 18.52  & 18.52  & 41.60 \\
dstLon           & 1.26  & 73.32  & -93.61 & 28.06  & 73.85 \\
\bottomrule
\end{tabular}
\end{table}

A closer examination of the categorical variables (Table~\ref{tab:categorical_summary}) reveals a highly skewed distribution across several dimensions. Although the dataset spans only three calendar days, attacks were concentrated on June 10 (6/10/25), reflecting operational or attacker-driven temporal biases. Source IP addresses are especially diverse, with 2,438 unique origins; yet the top origin (23.175.48.211) alone accounts for 18,561 events-nearly 15\% of the total-indicating that a small subset of hosts generates a disproportionately large volume of traffic. Autonomous system analysis shows a similar pattern, as the leading ASN (VDI-NETWORK) also contributes those same 18,561 events. Geographically, attacks emanate from 95 countries, though the United States dominates with over 38,600 events (around 31\%). On the destination side, only four countries host the honeypots, with India receiving 53,057 hits. In terms of protocol misuse, 13 distinct protocols appear, but Session Initiation Protocol (SIP) leads with 55,060 events-captured almost entirely by the SentryPeer honeypots, which logged the same volume-highlighting the prevalence of VoIP-related probing over more traditional services like Telnet or HTTP.

\begin{figure*}[t]
\centering
\begin{minipage}{0.75\linewidth}
  \centering
  \includegraphics[width=\linewidth]{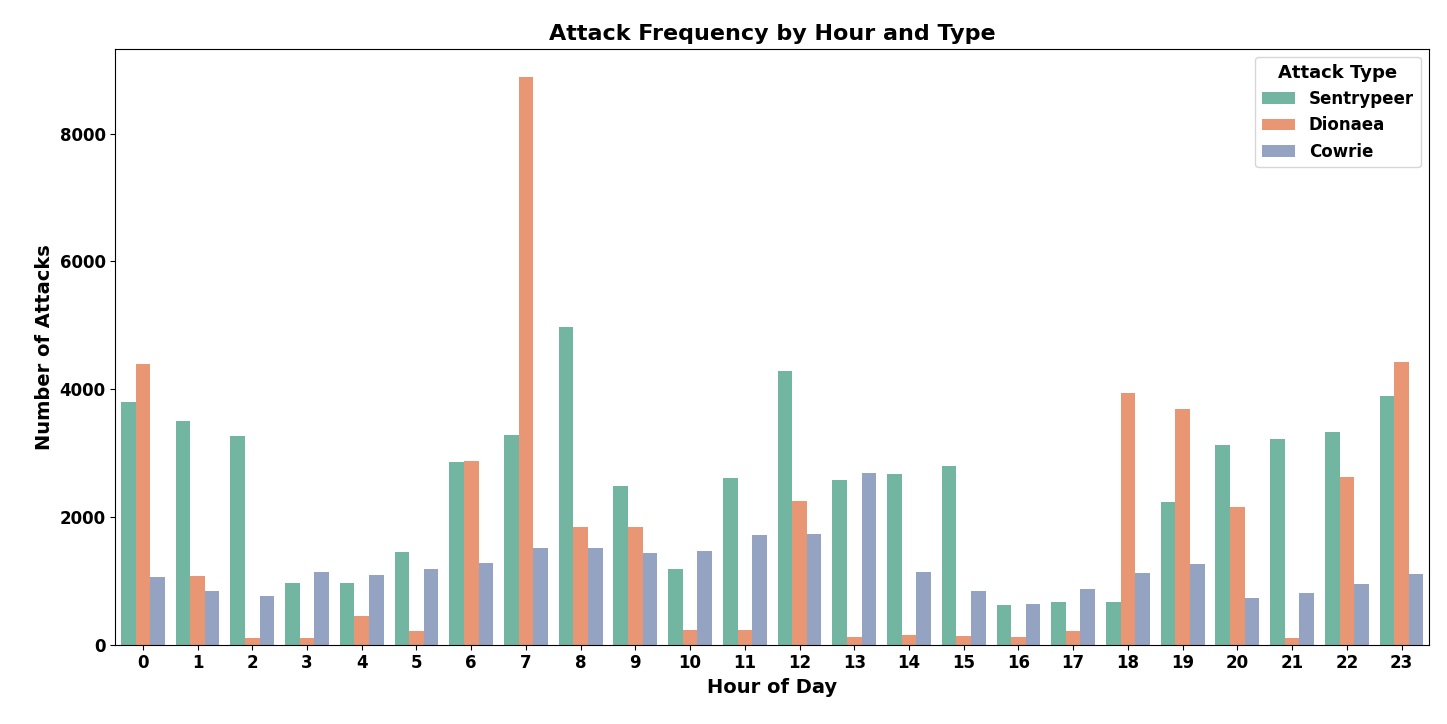}\\
  \vspace{0.2em}(a) 
\end{minipage}
\begin{minipage}{0.75\linewidth}
  \centering
  \includegraphics[width=\linewidth]{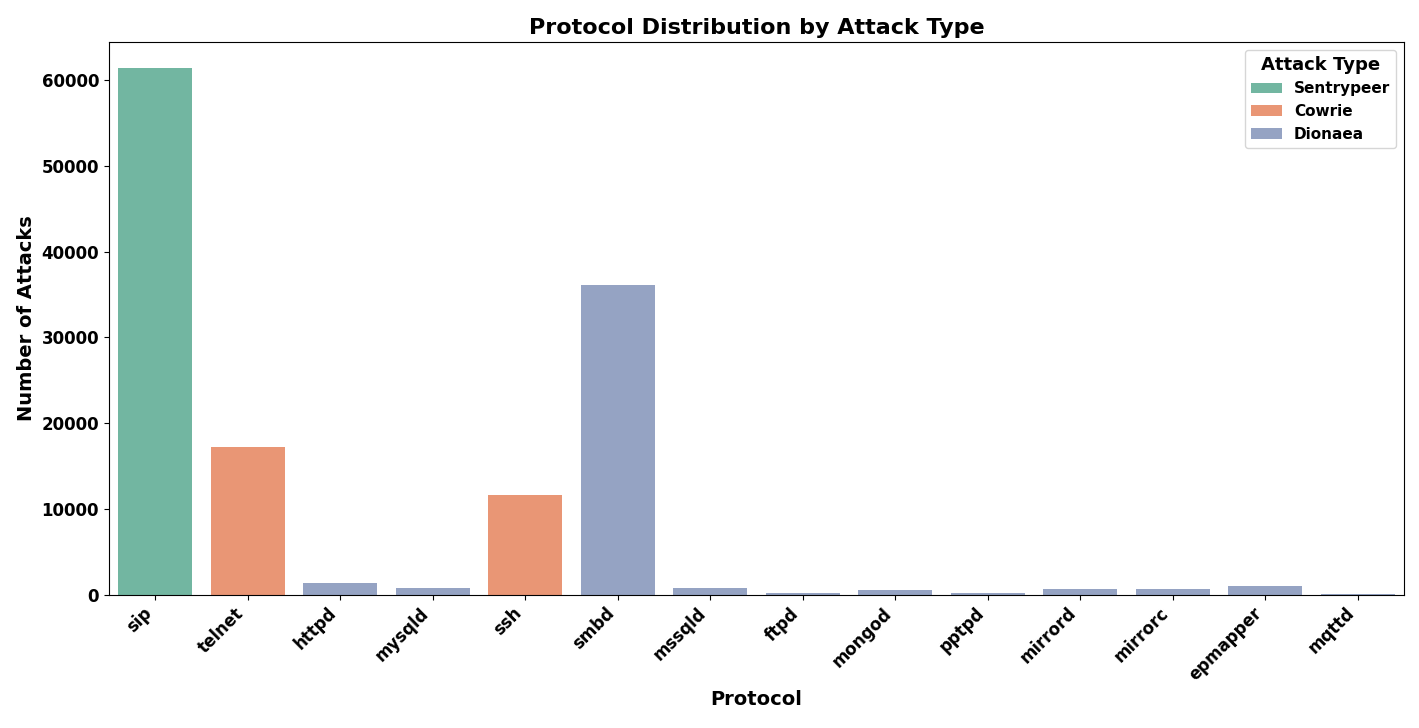}\\
  \vspace{0.2em}(b)
\end{minipage}
\caption{Visualizations of attack patterns. (a) Diurnal trends show peaks at 07:00 and 23:00 UTC. (b) Protocols abused by each honeypot type, with SIP notably exploited by SentryPeer.}
\label{fig:combined_attack_insights}
\end{figure*}

The numerical summaries (Table~\ref{tab:numerical_summary}) further elucidate the temporal and spatial dynamics of the attacks. Time-of-day analysis shows a mean occurrence at 11:41 UTC, but with a wide standard deviation of 7.19 hours, confirming nearly continuous probing around the clock. The interquartile range spans 06:00–18:00 UTC, with notable peaks observed at approximately 07:00 and 23:00 UTC (see Figure~\ref{fig:combined_attack_insights}(a)). Port statistics reveal a mean source port of 48,604, indicating the use of ephemeral ports for scanning, while destination ports exhibit extreme skew: a mean of 2,693 and a 75th percentile at 5,060 (SIP), signifying a concentration of attacks against VoIP endpoints.

These descriptive statistics underscore two fundamental characteristics of contemporary cyber-probing: extreme diversity in attacker origins (over 2,400 IPs across 95 countries) juxtaposed with concentration in top offenders (the single top IP and ASN generate ~15\% of events, and SIP dominates protocol usage), and a temporal persistence that spans all hours with well-defined peak periods. Such insights are of interest for designing adaptive defense strategies that account for both the long tail of sporadic scanning and the heavy hitters that drive the majority of intrusion attempts.

\begin{figure*}[t]
\vskip 0.1in
\centering
\begin{minipage}{\textwidth}
  \centering
  \includegraphics[width=0.9\linewidth]{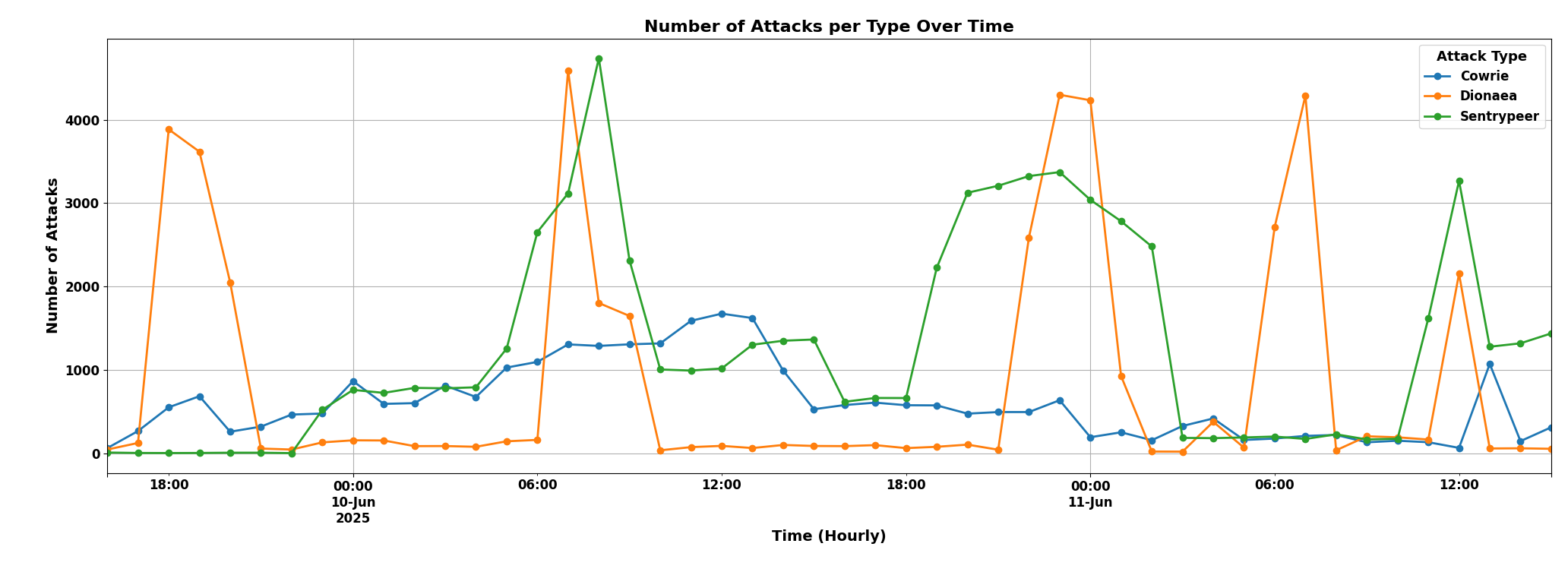}\\(a)
\end{minipage}
\vspace{1em}
\begin{minipage}{\textwidth}
  \centering
  \includegraphics[width=0.9\linewidth]{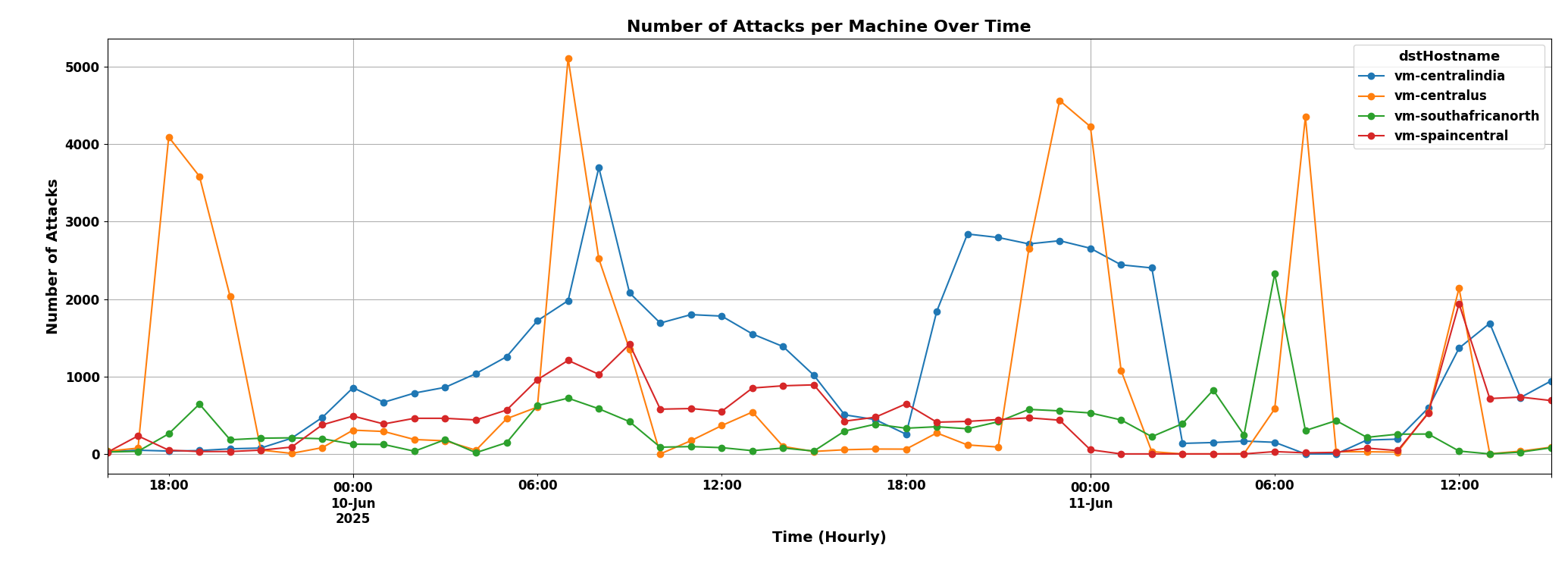}\\(b)
\end{minipage}
\caption{Temporal patterns of cyberattacks. (a) Evolution of attack types over time, highlighting peaks for Dionaea and SentryPeer. (b) Attack intensity across virtual machines, with \texttt{vm-centralus} being notably targeted.}
\label{fig:temporal_patterns}
\vskip -0.1in
\end{figure*}

Figure~\ref{fig:combined_attack_insights}(b) illustrates the distribution of protocols abused by each honeypot platform. Three distinct bars for each protocol show how SentryPeer, Cowrie, and Dionaea honeypots capture traffic on different services: (1) SentryPeer (blue bars) records an overwhelming concentration of attacks on SIP, with over 55,000 events, which reflects the platform specialization in VoIP-related probing and exploitation attempts; (2) Cowrie (orange bars) is dominated by Telnet traffic (approximately 29,000 events), consistent with its typical use as an SSH/Telnet honeypot that emulates legacy remote‐login services; (3) and Dionaea (green bars) exhibits a more diversified pattern exhibiting significant SIP and Telnet probing, although its highest volume (around 36,000 events) is against SMBD (Microsoft-DS/CIFS) endpoints. This underscores Dionaea focus on capturing malware propagation attempts targeting Windows file-sharing services.

Minor protocols such as HTTPD, MySQL, MSSQLD, MongoDB, PPTPD, Mirrord, MQTTP, and EPMAP collectively contribute only a few thousand events each, demonstrating that attackers overwhelmingly target the three primary vectors (SIP, Telnet, and SMBD) on these honeypots. The clear separation of protocol preferences by honeypot type highlights both attacker behavior and the inherent bias introduced by each honeypot service emulation.

Figure~\ref{fig:temporal_patterns}(a) plots the hourly evolution of total attacks broken down by honeypot type over the 72-hour window. Two pronounced sawtooth patterns emerge: first, sharp spikes in Dionaea (orange) around 18:00 UTC on June 9 and again just after 23:00 UTC on June 10; second, very high SentryPeer (green) peaks at approximately 07:00 UTC on both June 10 and June 11. These recurring rises and falls are not an artifact of attacker behavior alone but directly correspond to scheduled shutdowns and restarts of the virtual machines hosting those particular honeypots. Whenever a virtual machine goes offline, its event count abruptly drops to zero, only to climb again as it comes back online and resumes capturing probes. Cowrie (blue) shows a similar, though less dramatic, ripple pattern, reflecting its own maintenance cycles.

Figure~\ref{fig:temporal_patterns}(b) displays the aggregated attack volume per virtual machine hostname, irrespective of honeypot type. Here, the sawtooth is even more striking: each Virtual Machine-\texttt{vm-centralus}, \texttt{vm-centralindia}, \texttt{vm-southafricanorth}, and \texttt{vm-spaincentral}-exhibits intermittent valleys where attack counts drop to zero, followed by rapid recovery. Notably, \texttt{vm-centralus} (orange line) experiences the highest peak load (over 5,000 events in a single hour), emphasizing its attractiveness to attackers or its wider network exposure. The alternating high-frequency spikes and troughs map precisely to the operational schedule of the virtual machines, underscoring the importance of continuous monitoring and the potential for blind spots during maintenance windows.

\begin{figure*}[ht]
\centering
\includegraphics[width=0.75\linewidth]{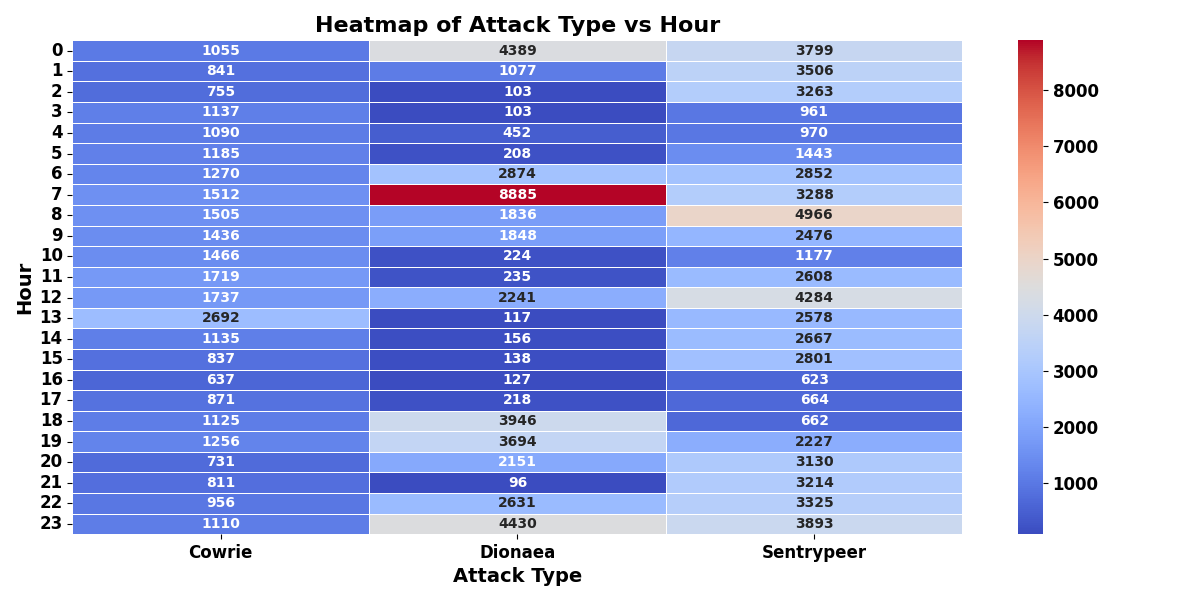}
\caption{Heatmap showing number of attacks by hour and attack type. Peak for Dionaea observed at 07:00 UTC.}
\label{fig:heatmap_hour_attack}
\end{figure*}

Figure~\ref{fig:heatmap_hour_attack} illustrates the diurnal intensity of honeypot engagements, with each cell encoding the number of logged attacks per hour for Cowrie, Dionaea, and SentryPeer. Dionaea exhibits a pronounced, narrow spike at 07:00 UTC (8,885 events), making it the single busiest hour across all platforms; this likely corresponds to coordinated probing campaigns against Windows file‐sharing services (SMBD). In contrast, SentryPeer shows an extended plateau of high activity between 06:00 and 13:00 UTC, peaking at 07:00 (3,121 events) and again at 08:00 UTC (4,740 events), reflecting sustained VoIP‐targeted scanning during standard business hours. Cowrie SSH/Telnet traffic is comparatively uniform, ranging from approximately 700 to 1,700 attacks per hour, with a modest high at 13:00 UTC (2,692 events), indicating around‐the‐clock interest in remote‐access services. All three honeypots record their lowest activity in the pre-dawn interval (02:00–04:00 UTC), suggesting a global lull in automated scanning or reduced attacker engagement during those hours. These distinct temporal signatures-short, intense bursts for Dionaea, broad daytime engagement for SentryPeer, and uniform probing for Cowrie-highlight the necessity of time‐aware defensive measures that can dynamically adjust to both rush‐hour surges and overnight lulls.

\begin{figure*}[ht]
\centering
\includegraphics[width=1\linewidth]{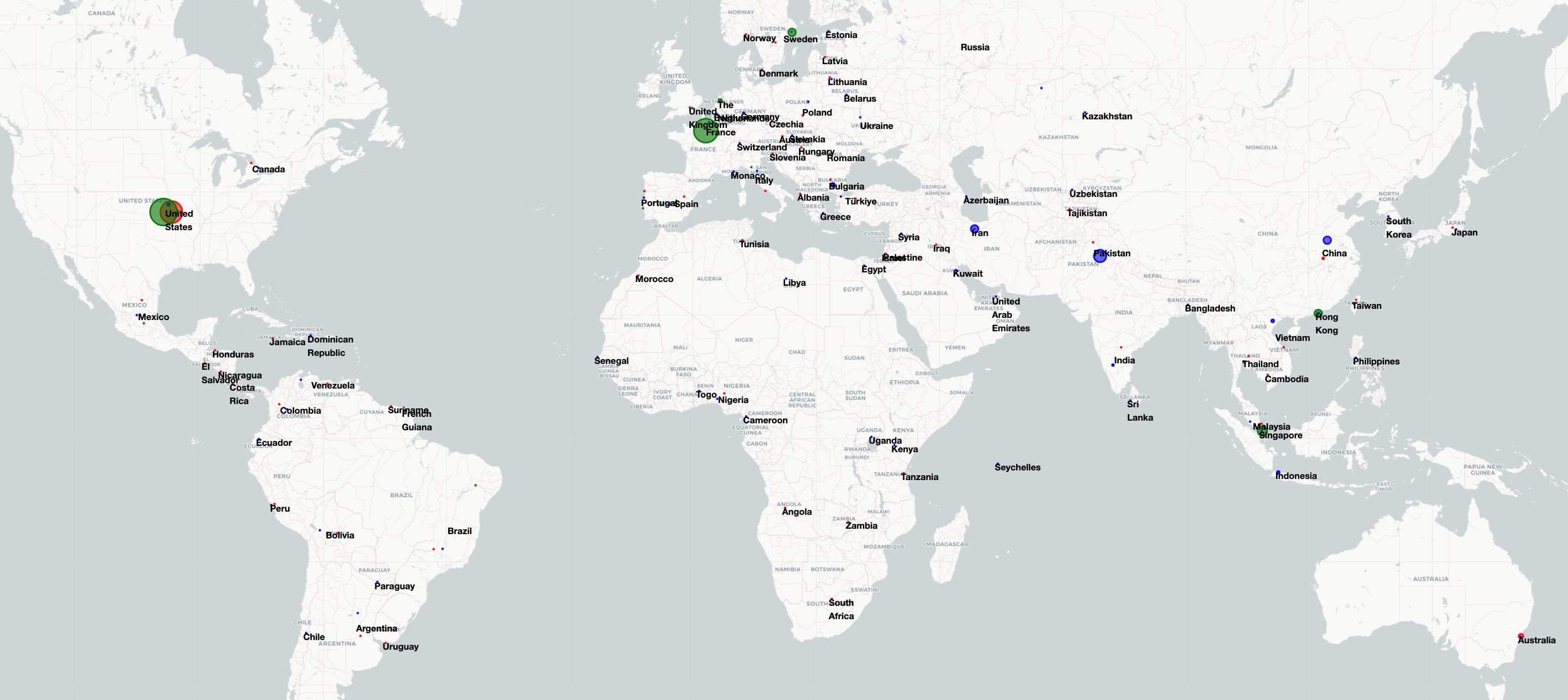}
\caption{Global distribution of cyberattacks by honeypot platform and country of origin. Circle size is proportional to the total number of attacks from each country, and color indicates the honeypot type: green (SentryPeer), red (Cowrie), and blue (Dionaea).}
\label{fig:world_map_attacks}
\end{figure*}

Figure~\ref{fig:world_map_attacks} maps the worldwide origins of all logged intrusion attempts, using circle diameters to encode the volume of attacks per country and hue to distinguish the honeypot platform involved. Green circles (SentryPeer) dominate regions such as the United States and parts of Southeast Asia, reflecting high SIP‐oriented probing. Red circles (Cowrie) are most prominent in North America and Western Europe, indicating concentrated Telnet/SSH scans. Blue circles (Dionaea) cluster around European nations-particularly France and the UK-signifying intensive SMB‐targeted activity. Smaller markers scattered across South America, Africa, and Oceania confirm a pervasive, though lower‐intensity, global scanning footprint. These spatial patterns suggest that while attackers operate worldwide, a handful of high‐connectivity regions generate the bulk of traffic, informing strategic placement and scaling of defensive honeypot deployments.

\section*{RECORDS AND STORAGE}

The dataset is organized into a root directory and two subfolders: \texttt{/RawData} and \texttt{/PreprocessedData}. Each file is described below to ensure reproducibility and ease of navigation.  

\subsection*{Root Directory Files}
The root directory contains files that document the dataset and define the preprocessing environment:
\begin{itemize}
    \item \texttt{DataPreprocessingNotebook.ipynb}: Jupyter notebook that loads, cleans, and transforms raw honeynet event data. It includes code for parsing JSON files, handling missing values, extracting features, and exporting CSV outputs.
    \item \texttt{README.md}: Plain text file containing general information about the dataset, its structure, and instructions for use.
    \item \texttt{requirements.txt}: Lists all Python dependencies required to run the preprocessing notebook, with specific version numbers to ensure reproducibility.
\end{itemize}

\subsection*{Raw Data Directory Files}
The \texttt{/RawData} folder contains the original, unprocessed logs of honeynet events in JSON format:
\begin{itemize}
    \item \texttt{RawHoneynetEventsBatch-0.json}: First batch of raw honeypot telemetry.
    \item \texttt{RawHoneynetEventsBatch-1.json}: Second batch of raw honeypot telemetry.
    \item \texttt{RawHoneynetEventsBatch-2.json}: Third and final batch of raw honeypot telemetry.
\end{itemize}

Each JSON file contains structured records of honeypot interactions, including timestamps, source IP addresses, target ports, and protocol types.

\subsection*{Preprocessed Data Directory Files}
The \texttt{/PreprocessedData} folder contains a single cleaned dataset derived from the raw JSON files:
\begin{itemize}
    \item \texttt{HoneyNetEvents\_Clean.csv}: Aggregated and preprocessed version of all raw batches, formatted as a CSV file. This dataset includes the extracted features defined in the preprocessing notebook.
\end{itemize}

The preprocessed CSV aggregates all three raw JSON batches and preserves a one-row-per-event convention. No features are computed that cannot be reproduced from the raw files and the provided notebook.

\section*{INSIGHTS AND NOTES}

This dataset presents several caveats and opportunities that researchers should consider. Maintenance-induced gaps due to virtual machine restarts create blind spots in time-series analysis and must be handled carefully, either by interpolation or explicit modeling as missing data. Each honeypot platform introduces inherent biases, with SentryPeer capturing SIP floods primarily from North America and Southeast Asia, Cowrie logging Telnet/SSH scans largely from Western Europe and the U.S., and Dionaea recording SMB exploitation around European nodes. Therefore, results should not be generalized without accounting for platform-specific characteristics. Attack activity is highly concentrated, with a small number of IPs generating a disproportionate share of traffic, highlighting the need for robust statistical methods suitable for heavy-tailed distributions. The four-region deployment also enables cross-regional analysis of attack patterns, though observed differences may partly reflect local Internet infrastructure rather than purely geographic intent. Beyond its primary use in threat analysis, the dataset offers opportunities for secondary applications such as benchmarking anomaly detection algorithms, studying temporal burstiness in event streams, and exploring protocol misuse in realistic environments, thereby extending its value beyond the scope of the original study.  Public IP addresses are retained as collected identifiers to support reproducibility; no additional personal data is included.

\section*{SOURCE CODE AND SCRIPTS}

Complete infrastructure deployment scripts, data collection tools, and analysis code are publicly available in the associated repository. The repository includes detailed documentation for replicating the experimental setup and extending the data collection methodology, including: infrastructure-as-code templates, honeypot configuration files, data processing pipelines, and analysis scripts \cite{Feito-Casares2025} \cite{Feito-Casares2025b}. All code is released under MIT License to facilitate academic and commercial reuse.


\section*{ACKNOWLEDGEMENTS AND INTERESTS}
This work was supported by the CyberFold project, funded by the European Union through the NextGenerationEU instrument (Recovery, Transformation, and Resilience Plan), and managed by Instituto Nacional de Ciberseguridad de España (INCIBE), under reference number ETD202300129. This work was also supported by the Autonomous Community of Madrid (ELLIS Madrid Node) and the Spanish Ministry of Science and Innovation through project PID2022-140786NB-C32/AEI/10.13039/501100011033 (LATENTIA).
\\
\text{\hspace{1em}} E.F.C. and I.G.T. curated and analyzed the data, designed the experimental infrastructure, and wrote the manuscript. J.L.R.A. supervised the research, provided methodological guidance, and reviewed the manuscript. All authors contributed to the conceptualization and review of the final manuscript.
\\
\text{\hspace{1em}} The authors have declared no conflicts of interest. We acknowledge the use of AI assistants, including large language models, to support the preparation of this work. These tools were used for tasks such as proofreading, improving clarity, and refining the structure of selected sections. All AI-generated suggestions were reviewed and approved by the authors to ensure academic integrity and accuracy.

\bibliographystyle{unsrtnat}
\bibliography{cas-refs}

\end{document}